\documentclass[%
reprint,
nofootinbib,
amsmath,amssymb,
aps,
prl,
twocolumn
]{revtex4-2}
\usepackage{url}% For URL to Particle Data Group in Bibtex
\usepackage{graphicx}% Include figure files
\usepackage{dcolumn}% Align table columns on decimal point
\usepackage{bm}% bold math
\usepackage[qm,braket]{qcircuit}
\usepackage{slashed}
\usepackage{hyperref}% add hypertext capabilities
\usepackage{csquotes}
\usepackage{relsize}
\usepackage{mathrsfs}
\usepackage{tikz}
\usepackage{bm}
\usepackage{euscript}

\newcommand{\mybar}[1]
        {\kern 0.6pt\overline{\kern -0.6pt#1\kern -0.6pt}\kern 0.6pt}

\newcommand{\be}{\begin{equation}}
\newcommand{\ee}{\end{equation}}
\newcommand\beq{\begin{eqnarray}}
\newcommand\eeq{\end{eqnarray}}
\newcommand\eqn[1]{\label{eq:#1}}
\newcommand\eq[1]{eq.~(\ref{eq:#1})}

\newcommand{\eV}{{\rm ~eV }}

\newcommand{\GeV}{{\rm ~GeV }}

\usepackage[section]{placeins}

\begin{document}

\preprint{INT-PUB-21-007 }

\title{Gravitational contributions to the electron $g$-factor}
\author{Andrew G.~Cohen} \email{acohen@ust.hk} \affiliation{%
  Institute for Advanced Study, Hong Kong University of Science and
  Technology, Hong Kong \ }%
\author{David B.~Kaplan} \email{dbkaplan@uw.edu} \affiliation{%
  Institute for Nuclear Theory, Box 351550, University of Washington,
  Seattle, WA 98195-1550 }

\begin{abstract}
In a previous paper, the authors with Ann Nelson proposed that the UV
  and IR applicability of effective quantum field theories should be
  constrained by requiring that strong gravitational effects are
  nowhere encountered in a theory's domain of validity
  [Phys. Rev. Lett. 82, 4971 (1999)]. The constraint was proposed to delineate the
  boundary beyond which conventional quantum field theory, viewed as
  an effective theory excluding quantum gravitational effects, might
  be expected to break down.  In this Letter we revisit this idea and
  show that quantum gravitational effects could lead to a deviation of
  size $(\alpha/2\pi)\sqrt{m_e/M_p}$ from the Standard Model
  calculation for the electron magnetic moment.  This is the same size
  as QED and hadronic uncertainties in the theory of $a_e$, and a
  little more than one order of magnitude smaller than both the dominant
  uncertainty in its Standard Model value
  arising from the accuracy with which  $\alpha$ is measured, as well as the
  experimental uncertainty in measurement of $a_e$.
  \end{abstract}
\maketitle

Effective field theory is a technique to exclude high energy
single-particle states from a theory's Hilbert space and construct a
Hamiltonian that accurately describes the physics of the remaining low
energy states.  Hermiticity within this restricted space ensures that
these low energy states do not evolve out of the restricted Hilbert
space with time.  This requires not only modifying interactions
between light degrees of freedom to account for the virtual effects of
the excluded particles, but also restricting the allowed particle
density; otherwise, an initial state with many widely separated low
energy particles could evolve to a denser state with sufficient energy
to produce the excluded heavy particles.  An example of such an
effective theory is a lattice version of QCD with lattice spacing $a$
providing a UV cutoff $\Lambda \sim \pi/a$ on the energy of
single-particle states. This lattice theory also limits the energy
density of multi-particle states to $\lesssim O(\Lambda^{4})$: for
fermions the density is restricted by the Pauli principle that allows
at most two particles per lattice site, while photons and gluons are
represented by compact link variables with similarly bounded energy
density\footnote{In a lattice Hamiltonian approach one would have to
  bound the magnitude of the electric field, such as by using compact
  variables for the electric field as in Ref.~\cite{Kaplan:2018vnj}.}.
Viewed as an effective field theory for the Standard Model with
$\Lambda = O(1\GeV)$ there are no states on the lattice which in the
full theory would evolve into propagating $W$ bosons, while the
effects of virtual $W$ bosons are accounted for by irrelevant
operators suppressed by powers of $M_W$.

It is conventional to treat quantum gravity similarly as a short
distance phenomenon that can be incorporated into effective theories
applicable at energies below the Planck Mass, $M_{p}$, by imposing a
cutoff $\Lambda\ll M_p$, with small residual effects accounted for by
irrelevant operators suppressed by powers of $M_p$. Weak gravitational
effects are included by allowing a classical background metric with
mild curvature.  However, when formulated in infinite volume, such an
effective theory unavoidably allows low energy particles to collapse
into what would be a black hole in General Relativity, even with an
arbitrarily low UV cutoff.  In the full theory including gravity this
would cause the formation of a spacetime singularity (or some quantum
smoothed version of one) hidden behind a horizon; this black hole
would emit Hawking radiation and shrink until the spacetime curvature
at the horizon reaches the Planck scale, where quantum gravitational
effects become important.  This superficially looks like many low
energy particles in lattice QCD being allowed to converge to create
propagating $W$ bosons, which is avoided by limiting the local energy
density; in the gravitational case, however, there is no local
constraint that can avoid black hole creation (which can occur at
arbitrarily low energy density), nor black hole evaporation (which can
occur by the emission of low energy particles).  While this discussion
treats black holes semiclassically, presumably black hole events can
also arise from quantum fluctuations in such a theory.  Therefore a
quantum field theory attempting to relegate the effects of quantum
gravity to irrelevant operators suppressed by powers of the Planck
mass cannot be fully consistent.  While perhaps it is possible to make
rigorous sense of a quantum field theory that treats quantum gravity
as a short distance phenomenon, it must be recognized that it does not
behave like a conventional healthy effective field theory: a theory
that excludes quantum gravity with corrections parametrized by
Newton's constant is very different than one that excludes quantum $W$
bosons with corrections parametrized by Fermi's constant.

Nevertheless, there is no denying that quantum field theories where
gravitational effects are parameterized by inverse powers of $M_p$
seem to work just fine.  Ref.~\cite{Cohen:1998zx} (CKN) argued that
such a theory makes sense provided it is formulated in finite volume,
providing an IR cutoff, with a particular constraint between the UV
and IR cutoffs. That work then analyzed whether one could
experimentally tell the difference between such a modified effective
theory and the conventional one.  CKN concluded that there were
important differences between the conventional and IR-modified
theories, with the size of gravitational effects at an energy scale
$E$ scaling with a lower power of $E/M_p$ than expected from
$M_p$-suppressed irrelevant operators.  Unfortunately, detection of
these exotic effects seemed out of reach of (then) current and future
experiments.  Recently, however, it was pointed out that part of the
CKN analysis included an error in estimating certain IR effects
\cite{Banks:2019arz}.  In this Letter we reanalyze the CKN theory
taking into account this correction. Our conclusion is that
gravitational corrections to the electron $(g-2)$ scale as
$\sqrt{m_e/M_p}$ instead of $(m_e/M_p)^{2/3}$ as stated in
Ref.~\cite{Cohen:1998zx}, and the deviation from the Standard Model is
tantalizingly close to the current experimental and theoretical errors
in the experimental determinations of $(g-2)$ and $\alpha$. Notably
these effects may be within reach of experiments in the foreseeable
future.

As we have argued, in a world with gravity it is not possible to
exclude gravitational effects in an effective theory in infinite
volume by means of local constraints on the fields.  The CKN starting
point was to discard the notion that an effective theory can be used
with a given UV cutoff in an arbitrarily large volume.  By requiring
that the theory not evolve states into black holes CKN arrived at a
simple constraint between the cutoff $\Lambda$ and size $L$ of such a
theory:
\beq \Lambda^2 L \lesssim M_p\ .  \eqn{UVIR} \eeq
This constraint relating UV and IR cutoffs implies that the usual
procedure of sending both $\Lambda$ and $L$ to infinity is not
possible, and consequently unavoidable uncertainties in perturbative
quantum field theoretic calculations remain even at arbitrarily small
coupling. As discussed in \cite{Cohen:1998zx}, this bound is more
stringent than one might argue based on simple holographic
considerations.

While this argument makes no presumptions regarding the cosmological
constant problem, intriguingly
the Standard Model with IR cutoff $L$ set to the current horizon size
has a UV cutoff from \eq{UVIR} of $\Lambda\sim 10^{-3}\eV$ and a
resulting vacuum energy consistent with the observed cosmological
acceleration.  There have been various cosmological models proposed
exploiting a UV/IR constraint to explain the dark energy, such as
Ref.~\cite{li2004model}, but while they are intriguing, it would be
more compelling to find evidence for the UV/IR constraint in
terrestrial experiments.  Since the most accurate tests of quantum
field theory are the $(g-2)$ of the electron and muon, it makes sense
to consider what happens to the calculation of these at one loop
\footnote{We note that there have been other suggestions in recent
  years for terrestrial experiments to find novel gravitational
  effects arising from an IR/UV correspondence, although they do not
  appear to be related to what we discuss here \cite{
    chou2017holometer,chou2017interferometric,
    verlinde2019observational,Verlinde_2020,zurek2020vacuum}.}.

The conventional 1-loop result for $a \equiv (g-2)/2$ for a lepton of mass $m$  is
% ,
\beq
a(L,\Lambda) &=& \frac{4
  \alpha m^2}{\pi}\int_0^1 dx\int_0^{(1-x)}dy \,(x+y)(1-x-y)\cr
&&\qquad\times \int_0^\infty dk\, \frac{k^3}{(k^2 + \Delta)^3}\ ,
\eeq
with $ \Delta = m^2(x+y)^2$.  Although the $k$ integral is finite, the
CKN prescription evaluates it with an IR cutoff
$k_\text{min} = 2\pi/L$ and a UV cutoff $k_\text{max} = \Lambda$.  The
result is
\beq
a(L,\Lambda)   =\left(\frac{\alpha}{2\pi}\right)\left(1-
  \frac{\pi^2}{mL} -\frac{m^2}{3 \Lambda^2} + \cdots\right)\ ,
\eeq
where the first term is Schwinger's classic result, and we have kept
only the leading\footnote{Ref.~\cite{Cohen:1998zx} erred in finding a
  leading $L$ dependence of $1/L^2$ instead of $1/L$; however, it
  should be noted that that reference advocates a different
  interpretation to the UV/IR constraint \eq{UVIR} and predicts a
  negligible effect on $(g-2)$. The $1/L$ behavior was noted in
  \cite{davoudi2014finite}. } dependence on $L$ and $\Lambda$.  The
UV/IR constraint in \eq{UVIR} restricts us to a 1-parameter set of
effective theories: choosing a large UV cutoff to minimize the
sensitivity to UV physics necessitates taking a low IR cutoff which
increases sensitivity to large extended states.  Our best strategy to
reduce sensitivity to strong gravitational effects is to minimize the
combined UV and IR corrections in the above result.  Minimization is
achieved by choosing cutoffs
\beq
\Lambda \sim m \left(\frac{M_p}{ m}\right)^{1/4}\ ,\qquad
L \sim \frac{1}{m}\left(\frac{ M_p}{m}\right)^{1/2}\ .
\eeq
For $m=m_e$, this results in $\Lambda \simeq 200\GeV$ and
$L\simeq 6$~cm, and yields the minimum deviation from the Schwinger
one-loop result:
\beq
a -
\frac{\alpha}{2\pi} \sim \frac{\alpha}{2\pi}
\sqrt{\frac{m}{M_p}}=\begin{cases} 10^{-14} & m=m_e\\ 10^{-13} & m =
  m_\mu\end{cases} .  \eqn{dev}
\eqn{dev}
\eeq
We emphasize that we are not pretending to  compute corrections to
$a_e$ from quantum gravity---but only the intrinsic uncertainty to the
calculation in an effective theory that excludes gravity. 

These deviations from the conventional result are experimentally
interesting for the electron.  Current experimental and theoretical
values for $a_e$ are quoted in
Ref.~\cite{Morel:2020dww,parker2018measurement,aoyama2019theory} as
\beq
a^\text{th}_e\text{(Rb)} &=&
% 1\ 159\ 652\ 182.037\ (720)(11)(12) \times 10^{-12}\ ,\cr   ** old value in Aoyama replaced by new value in Morel.
1\ 159\ 652\ 180.252\ (95)(11)(12) \times 10^{-12}\ ,\cr
a^\text{th}_e\text{(Cs)} &=& 1\ 159\ 652\ 181.606\ (229)(11)(12) \times  10^{-12}\ ,
\nonumber
\eeq
where the two values depend on whether one uses the measurement of
$\alpha$ from atomic experiments using Rb
\cite{PhysRevLett.106.080801} or Cs \cite{parker2018measurement}
respectively.  The three errors quoted are due to uncertainties in the
measurement of $\alpha$, numerical evaluation of tenth order QED, and
hadronic contributions respectively \footnote{The theoretical
  extraction of $\alpha$ from atomic experiments also incurs
  uncertainties $\delta \alpha/\alpha$ from the truncation of the
  Hilbert space, at the level one part in
  $\alpha/2\pi\sqrt{m_e/M_p}\sim 10^{-14}$.  However, this is much
  smaller than the current quoted error of one part in $10^{-10}$.
  The enhanced sensitivity of $(g-2)$ to gravitational effects stems
  from the fact that leading contributions to it are radiative
  corrections.}.

Currently the best experimental
measurement of $a_e$ is given by \cite{PhysRevLett.100.120801}
\beq
a^\text{expt}_e = 1\ 159\ 652\ 180.73\ (28) \times 10^{-12}\ .  
\eeq
We see that the hadronic and QED uncertainties in the theoretical
value are at the $10^{-14}$ level, roughly the same size as the
deviation we find in \eq{dev}, while the experimental uncertainties in
$\alpha$ and $a_e$ are about one order of magnitude larger.
Therefore, while observing the $ (\alpha/2\pi)\sqrt{m_e/M_p}$
discrepancy would require significant progress in both experiment and
theory, such progress is entirely possible.  The same cannot be said
for the muon, given that current experimental and theoretical
uncertainties in $a_\mu$ are at the $10^{-10}$ level
\cite{Aoyama:2020ynm}, three orders of magnitude larger than the
expected $ (\alpha/2\pi)\sqrt{m_\mu/M_p}$ deviation.

Our truncation of the Hilbert space through simultaneous UV and IR
cutoffs on (Euclidean) momenta is admittedly crude but has the virtues
of respecting manifest Lorentz invariance and allowing for a simple
power counting. It should be noted that Banks and Draper in Ref.~\cite{Banks:2019arz} 
 reinterpret the bound \eq{UVIR} in terms of a scale-dependent modification in the density of states and  arrive at a much smaller (and unmeasurable) deviation for $(g-2)$ than we obtain in \eq{dev}.  If such an approach can be made precise it would be most welcome, but we
do not currently know how to implement a scale-dependent
depletion of states that shares the above virtues.

\bigskip

This work is dedicated to the memory of Ann Nelson.  We thank Tom
Banks, Howard Georgi, Martin Hoferichter, and Dam Son for useful
communications.  This research was supported in part by DOE Grant
No. DE-FG02-00ER41132, and by Grant No AoE/P-404/18-3 of the Research
Grants Council of Hong Kong S.A.R.

%%%%%%%%%%%%%%%%%%%%%%%%%% 

\bibliography {BH}
\bibliographystyle{unsrt}

\end{document}